# Passive Cervical Spine Ligaments Provide Stability during Head Impacts in Vivo

Calvin Kuo[1*], Jodie Sheffels[2], Michael Fanton[1], Ina Bianca Yu[2], Rosa Hamalainen[1], David Camarillo[2]

[1]Mechanical Engineering Department, Stanford University, 443 Via Ortega, Room 202, Stanford CA 94305
[2]Bioengineering Department, Stanford University, 443 Via Ortega, Room 202, Stanford CA 94305

*Corresponding Author:
Name: Calvin Kuo
Address: 443 Via Ortega, Room 202, Stanford CA 94305
E-mail: calvin.kuo@ubc.ca
Telephone: (778) 680-0219

**Abstract**
It has long been suggested that neck muscle strength and anticipatory cocontraction can decrease head motions during head impacts. Here, we quantify the relative angular impulse contributions of neck soft tissue to head stabilization using a musculoskeletal model with Hill-type muscles and rate-dependent ligaments. We simulated sagittal extension and lateral flexion mild experimental head impacts performed on 10 subjects with relaxed or cocontracted muscles, and median American football head impacts. We estimated angular impulses from active muscle, passive muscle, and ligaments during head impact acceleration and deceleration phases. During the acceleration phase, active musculature produced resistive angular impulses that were 30% of the impact angular impulse in experimental impacts with cocontracted muscles. This was reduced below 20% in football impacts. During the deceleration phase, active musculature stabilized the head with 50% of the impact angular impulse in experimental impacts with cocontracted muscles. However, passive ligaments provided greater stabilizing angular impulses in football impacts. The redistribution of stabilizing angular impulses results from ligament and muscle dependence on lengthening rate, where ligaments stiffen substantially compared to active muscle at high lengthening rates. Thus, ligaments provide relatively greater deceleration impulses in these impacts, which limits the effectiveness of muscle strengthening or anticipated activations.

**Introduction and Background**

Anecdotally, muscle strength and muscle cocontraction in anticipation of an impact are thought to reduce head motions during a head impact, which in turn is thought to reduce risk of brain injury. Clinicians, athletic trainers, and coaches have thus been recommending neck muscle strength training and field awareness exercises for contact sports athletes to protect themselves on the field (1).

Research studies have been less conclusive about the correlation between muscle strength or anticipatory muscle cocontraction with reduction in head motion. Laboratory studies applying low severity external loads to human subjects have demonstrated that anticipated muscle cocontraction significantly reduces head motion following impacts (2). However, impact forces in these studies are on the order of maximal neck muscle isometric force production, and it is unclear how effective neck muscles are at resisting more severe impact forces.

One prospective field study found a correlation between neck muscle strength and reduced brain injury incidence in high school athletes (3). This represented the first field evidence that a correlation exists between muscle strength and reduction of brain injury, which the authors postulated was due to a reduction in head motion following impacts. However, the study neglected to account for factors such as athlete mass which have also been implicated in head impact kinematics (4). Other prospective field studies have also failed to identify similar correlations between neck muscle strength or anticipatory preparation and brain injury risk or head kinematic severities (5, 6).

Researchers have also investigated the effect of neck strengthening regimens on head motion, which are more useful for suggesting preventative measures for individuals (7, 8). These studies have found that while strengthening regimens can increase neck strength, there was little evidence that such techniques reduced head motions. However, many of these studies were limited by relatively small subject populations, and a large prospective study relating neck strengthening exercises and reduction of head motion following an impact is a current gap in the literature.

The studies discussed thus far have focused on drawing correlations between neck muscle strength or anticipated cocontraction with decreases in head motion following an impact. To determine an anatomical mechanism by which neck muscle strength or anticipated cocontraction might reduce head motions following an impact, a more controlled or model-based approach is required.

Recently, several researchers have sought to uncover this mechanism by simulating head impacts while varying neck muscle strength or activity (9–11). However, researchers have reached varying conclusions due to differences in study design, such as choice of model (finite element, anthropomorphic test dummy surrogate, rigid linkage model) or choice of impact conditions (severity, direction, impact surface). Furthermore, these studies have focused primarily on the effect of active muscles, and typically neglect the role of other soft tissue in stabilizing the head.

Thus, the role of neck muscles in head impacts remains an active area of research. In this study, we approach this problem by quantifying the moment and angular impulse contributions of active musculature, passive muscle structure, ligaments, and external loads to study their relative roles during head impacts using musculoskeletal head and neck models. We simulated experimental head motions performed in 10 subjects during mild head impacts and extrapolated to median American football head impacts. The moment production of individual elements are dictated by their constitutive material models, which shed light on the underlying biomechanical mechanisms of head stabilization during impacts.

**Results**

To determine the relative contributions of muscle activity, passive muscle structures, and passive ligaments during head stabilization, we modified a previously developed OpenSim head and neck musculoskeletal model (12, 13). We performed forward dynamics simulations of previously published mild experimental head impacts in 10 subjects (4). In the experiments, loads were applied to induce head motion in two directions (sagittal plane extension and coronal plane lateral flexion towards the non-dominant side) and with two muscle activity conditions. For the first muscle activity condition, we instructed subjects to minimally activate neck muscles while maintaining an upright posture (gravity balance), which we consider a relaxed condition. For the second muscle activity condition, we instructed subjects to maximally activate muscles while maintaining an upright posture, which we consider a cocontracted muscle condition.

Subject-specific models were generated for these simulations scaled to subject height, mass, and isometric strength in sagittal plane flexion and coronal plane lateral flexion. In each simulation, motion was restricted to the primary motion plane (sagittal plane or coronal plane). From the simulations, we determined the relative moments and angular impulses responsible for head stabilization. We also simulated median severity American football head impacts with cocontracted muscles in the 10 subject-specific models using extrapolated force profiles to observe differences in relative moment and angular impulse contributions.

*Experimental Angular Impulse and Moment Contributions*
To demonstrate contributions to stabilization during mild experimental head impacts, we first show relative angular impulse contributions of the external force, gravity, all muscles (active component), all muscles (passive component), and all ligaments during acceleration and deceleration phases (Figure 1). Contributions were normalized by the total angular impulse provided by the external load and aggregated over all 10 subjects for each condition. Angular impulses were obtained by integrating moments over periods of acceleration and deceleration.

In all conditions, the acceleration phase was dominated by the angular impulse provided by the external load. When muscles were relaxed, there were no substantial resistive angular impulses provided by any soft tissue structures during the acceleration phase. However, when the muscles were cocontracted, the active muscles provided resistive angular impulses that were 30% of the total impact angular impulse.

During the deceleration phase, soft tissue structures were responsible for providing angular impulses that stabilized the head. When subjects relaxed neck muscles, angular impulse were dominated by the ligaments and passive muscle in both sagittal extension and coronal lateral flexion. When muscles were cocontracted, angular impulse contributions were dominated by the active muscle.

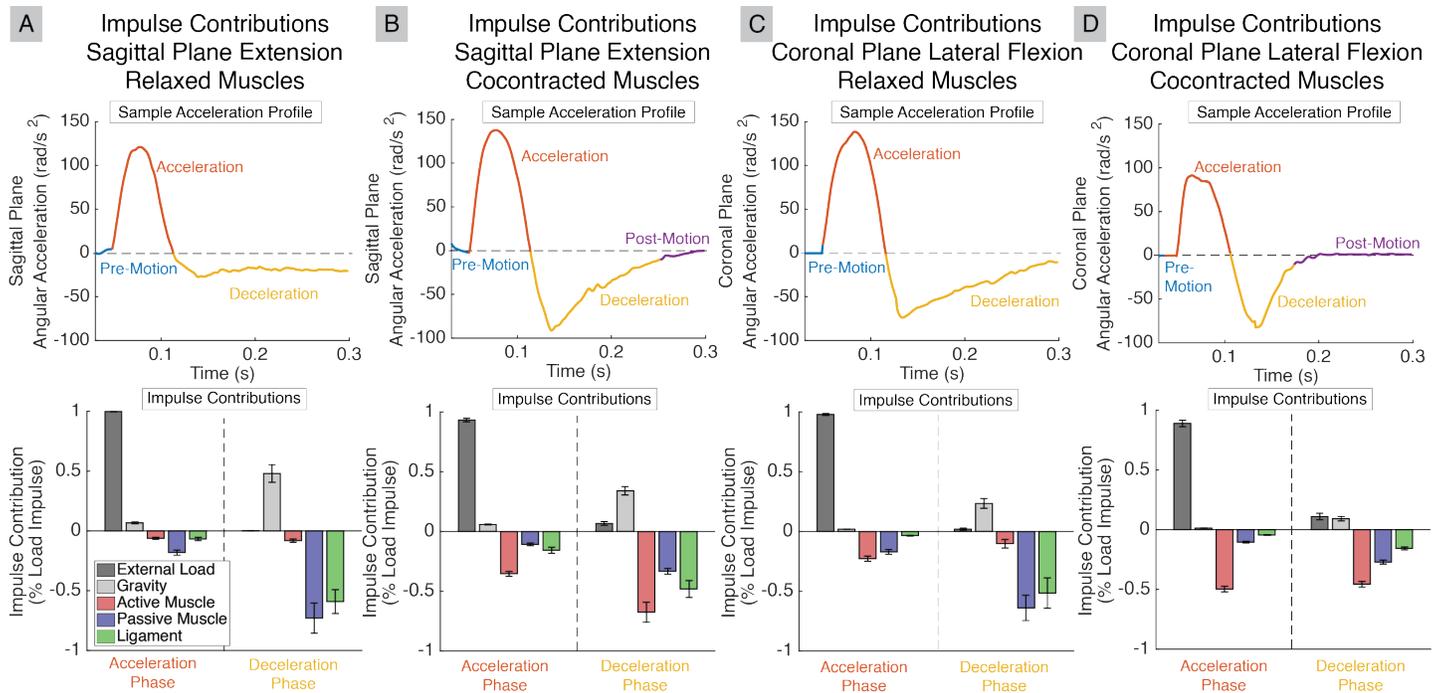

**Figure 1: Angular Impulse Contributions during Simulated Mild Head Impacts:** Impacts were divided into an acceleration and deceleration phase based on the planar angular acceleration time history (positive acceleration is sagittal extension of lateral flexion towards the non-dominant side). We included a short 30ms of zero force pre-load to allow the model to reach a balanced steady-state (pre-motion). In some conditions, head stabilization was achieved before the end of the 300ms impact simulation, resulting in a post-motion period with low angular accelerations ($|\alpha|<10$ rad/s$^2$). Impulse contributions during acceleration and deceleration phases were computed for external load, gravity, all muscles (active component), all muscles (passive component), and all ligaments for each condition and aggregated over all subjects. Angular impulse contributions were also normalized by total angular impulse produced by the external force. When neck muscles were relaxed, ligaments and passive muscles had the largest angular impulse contribution in deceleration for (A) sagittal extension and (C) coronal lateral flexion respectively. When neck muscles were cocontracted, the active muscle had the largest angular impulse contribution in deceleration for both (B) sagittal extension and (D) coronal lateral flexion. In all cases, external load provided the angular impulse that accelerated the head, with active muscles providing over 30% resistive angular impulse during cocontracted muscle cases.

We next show a breakdown of moments generated over time by external load, gravity, muscle groups (active component), muscle groups (passive component), and ligament groups (Figure 2). These were produced for a single subject with relaxed neck musculature in sagittal extension and coronal lateral flexion to pinpoint the structures most responsible for head stabilization, and to demonstrate the time evolution of moment generation.

In sagittal extension, both active and passive components of the hyoid muscles and sternocleidomastoid (SCM) muscles had large negative moment contributions resisting impact motion. Annulus Fibrosous (AF) fibers produced significantly greater negative moments among the ligaments. While both AF fibers and active SCM muscles produced similar negative moments, there were some muscles, such as the Splenius Capitis, providing positive moments that reduced the overall contribution from the active muscles. These positive moments were due to the initial activation that was necessary to maintain an upright posture (gravity balance) and was held through the duration of the impact.

In coronal lateral flexion, the SCM produced the greatest negative moment of the muscles (active and passive components), with substantial contributions from the passive component of the Trapezius and Scalene muscles as well. Of the ligaments, the joint capsules provided the greatest negative moments, with minor contributions from the AF fibers and the Ligamentum Flavum.

*Extrapolated Median Football Impact*
Finally, we show results from the median severity American football head impacts simulated using an extrapolated force profile aggregated over the 10 subjects (Figure 3). The force profile was generated to produce median American football kinematics over the 10 subject-specific models with cocontracted muscle activations. Median kinematics were taken from a previous exposure study of American football athletes wearing instrumented mouthguards over several games (14, 15). Simulated kinematics were within 20% of the median values in both sagittal extension and coronal lateral flexion.

Angular impulse analysis (similar to Figure 1) shows a decrease in resistive active muscle angular impulse contribution during the acceleration phase in both sagittal plane extension and coronal plane lateral flexion directions. While active muscles produced a resistive angular impulse that was 30% of the total impact angular impulse in mild experimental loads, this is reduced to below 20% in median American football head impacts.

During the deceleration phase, there is a relative decrease in active muscle angular impulse contribution despite using cocontracted muscle activations. In sagittal extension, active muscles in fact provide a positive angular impulse, whereas the ligaments now provide the largest resistive angular impulse contribution. The positive angular impulse is likely due to the SCM providing an extension moment at large extension angles as reported previously (12). In coronal lateral flexion, both the passive ligaments and passive muscle structures provide larger resistive angular impulse contributions than the active muscles.

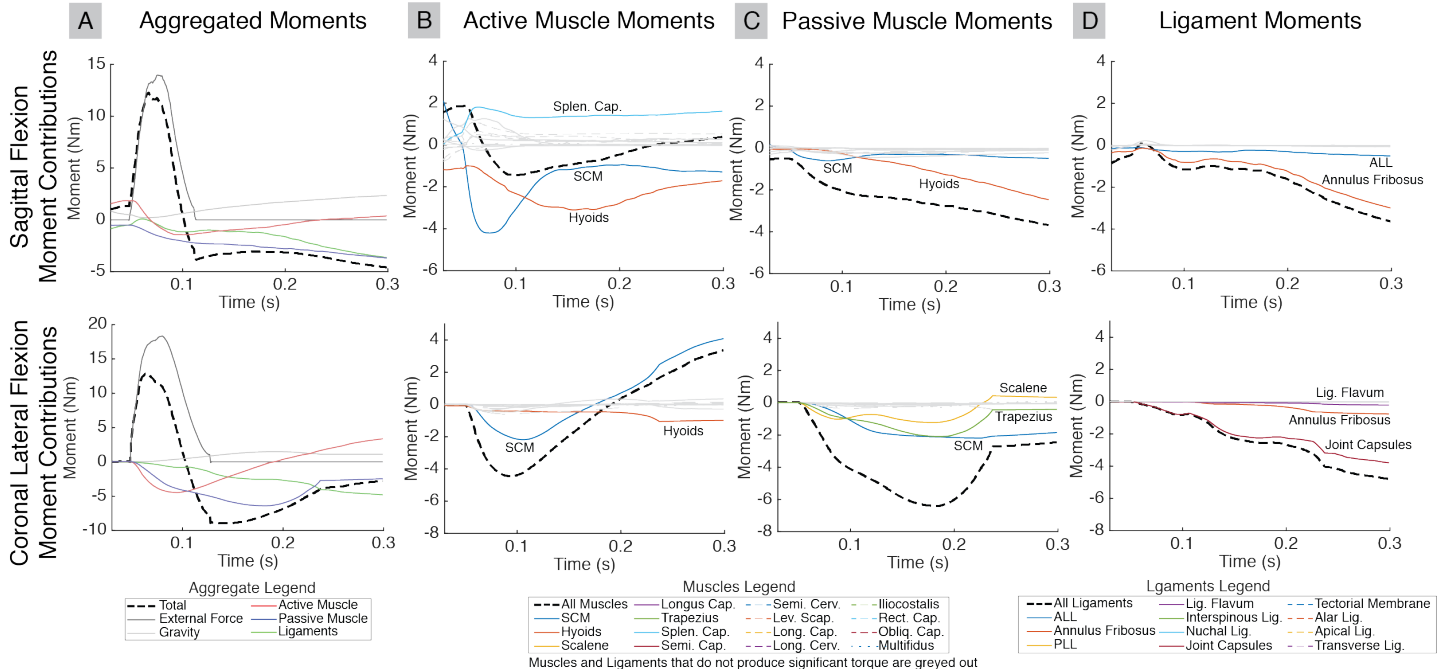

**Figure 2: Moments Produced by Model Forces Demonstrate Important Structures for Stabilization:** Moments over the impact period from force elements in simulation are presented (positive moments are moments in sagittal extension or lateral flexion to the non-dominant side). We show samples from a sagittal extension and coronal lateral flexion simulation with relaxed neck muscle activations. (A) Moments from external load, gravity, aggregated active muscle, passive muscle, and ligaments are shown first to demonstrate how moments from each group change with time. Moments from (B) active muscle component and (C) passive muscle component show that the SCM produces large negative moments resisting impact motion in both directions. Moments from (D) ligaments show that AF fibers and Capsular Ligaments dominate in sagittal extension and coronal lateral flexion respectively.

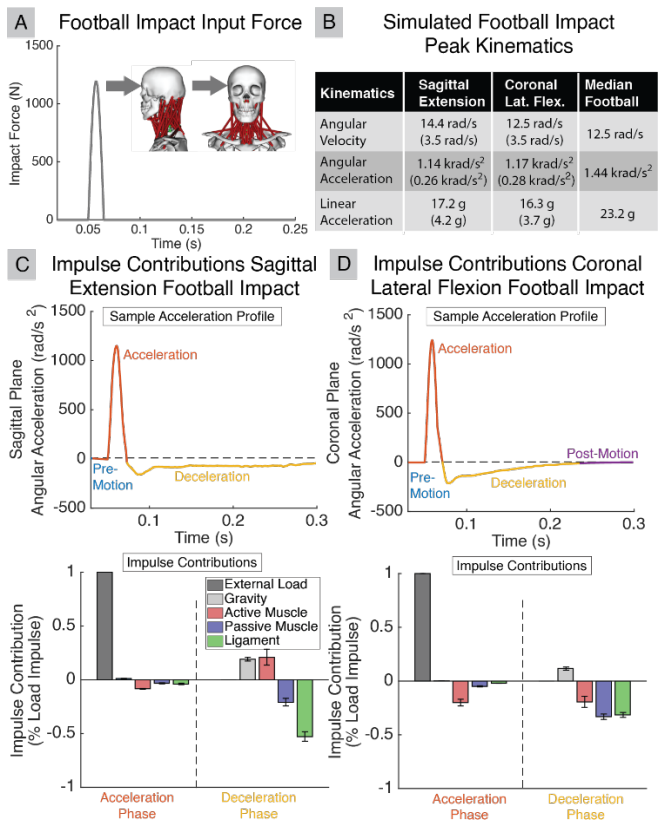

**Figure 3: Extrapolated American Football Median Impact Shows Change in Stabilization Contributions:** (A) We applied a simulated American football impact force in sagittal extension and coronal lateral flexion to our strongest subject with cocontracted neck muscle activations. (B) Simulation kinematics were within 20% of median American football impact kinematics. Angular impulse contributions in (C) sagittal extension and (D) coronal lateral flexion show that the active muscle has less relative contribution than in the experimental trials. In sagittal extension, the ligaments provide the most angular impulse in deceleration.

**Discussion**

In this study, we estimate the relative angular impulse contribution of active muscle, passive muscle, and passive ligament structures to head stabilization during experimental mild head impacts and extrapolated median American football head impacts using a musculoskeletal OpenSim model. While many previous studies have focused on the ability of neck musculature to resist head impacts and stabilize motion, we sought to quantify the contributions of the various cervical spine tissue structures (including the active and passive components of muscles).

We have demonstrated that in experimental mild head impacts, cocontraction of neck muscles resulted in large stabilization moments from the active component of the muscles, in particular the sternocleidomastoid (SCM) muscle. As reported in our previous study where experimental data were analyzed in more detail (4), head kinematics in impacts with cocontracted muscles were lower, likely due to the significantly increased stabilization contribution from the muscle activity. This also corroborates other similar experimental studies applying mild loads to subjects (2). However, when we extrapolate to median severity American football impacts, active muscle contributions decrease relative to other elements (ligaments and passive muscles). In sagittal extension, one of the more common impact scenarios in American football (14), the passive ligaments provide the most angular impulse stabilizing the head.

This indicates that the relative distribution of loads in the cervical spine among muscles and passive structures is dependent on impact direction and severity. In coronal lateral flexion, it has been previously reported that muscle moment arms for neck lateral flexors are larger than for neck sagittal flexors (12). The SCM, which contributed the most muscle moments and angular impulses in our study, had a nearly five times larger moment arm in coronal lateral flexion than sagittal flexion (12). Thus, the muscle angular impulse contributions remain relatively large in lateral flexion.

For severity dependence, we focus on the constitutive models for the muscle (Hill-Type) and ligaments (Figure 4). At higher severities, the lengthening rates of both the muscles and ligaments increase substantially. Previous studies of whiplash simulations report that ligaments can experiences over 1000mm/s lengthening rates (16–18). While the velocity relationship for Hill-Type active muscle plateaus at a relatively low lengthening rate (one optimal fiber length per second), which we achieved during our mild experimental head impacts, it has been suggested that the ligaments have a log-stiffening behavior with respect to lengthening rate (18–20). Thus, with increased impact severity, the active muscles did not produce substantially greater moments than the mild experimental head impacts. However, due to the strong ligament dependence on lengthening rate, the ligament moments scaled with impact severity.

While we were able to simulate experimental mild head impacts using an OpenSim model and analyze muscle and ligament contributions to stability, this study has a number of limitations. First, our findings are dependent on our chosen constitutive muscle and ligament models, and the properties for each muscle and ligament. However, we note that the Hill-Type muscle model is a standard model used in countless musculoskeletal modeling platforms to model all aspects of the human body (19, 21, 22). The ligament model, particularly the force-length relationship, is also well established in the literature and used in many finite element models (19, 20). We note however that muscles typically are not evaluated at extremely high lengthening rates, and while force-velocity scaling plateaus in the current Hill-Type muscle model implementation, the scaling may continue to increase at extremely high rates.

The OpenSim model also only includes structures representing the skeletal structure, ligaments, and muscles. However, the cervical spine contains other structures (such as the trachea) that could also contribute to angular impulses during impact . We included only muscles and ligaments as they are the most ubiquitous in the cervical spine, but other structures should be properly modeled and analyzed further. We note, however, that muscles represent the only active structure in the cervical spine, so the addition of additional tissue structures will only add to impulses from passive components that are not affected by muscle strengthening or pre-activation.

For the OpenSim simulations, we chose to fix the torso to ground (Methods). We did this because in the experimental mild head impact study, we attempted to fix the torso by seating subjects in a rigid back chair (4). However, as was reported in the previous study, there was some torso motion, particularly in the coronal lateral flexion trials. We note that linear accelerations had the largest errors between OpenSim simulations and experimental head impacts (Supplementary Information D), and it is possible this was because of the torso motion. However, we note that our other analyses focused on angular impulses and moments (angular metrics).

For coronal lateral flexion experiments, we had to constrain the OpenSim model to planar rotations by locking out-of-plane joints. It has been previously reported and it was observed in the experimental dataset that there is significant coupling between coronal lateral flexion and axial rotations (23). While these out-of-plane rotations were observed experimentally, we previously reported that the in-plane motion dominated the head motion response experimentally (4).

Other methods for computing intervertebral moments were previously presented for a finite element model in whiplash scenarios (24). These methods involved computing moments in a planar cross-section, and were thus sensitive to choice of cross-section and did not account for moments from forces not captured within the cross-section (specifically intervertebral discs as is noted in the study) (24).

In conclusion, we have quantified the relative contribution of muscles and ligaments to head stabilization during impacts. While there is no doubt the muscles play a role in stabilizing the head, their relative contribution depends on impact severity and direction and is not always greatest. Thus, claims that neck muscle strengthening and anticipatory cocontraction reduce head motions following impact must be made cautiously, and should be compared against other factors such as head and neck orientation (4, 25). Cervical spine ligaments are well studied in whiplash injuries and automotive crashes, and despite playing a large role in head and neck stability, they have received relatively little attention in head impact biomechanics.

Our OpenSim model and analysis represent a step towards uncovering the nuanced roles of the cervical spine structures in head impacts. However, further improvements can and should be made. To facilitate collaborative efforts and dissemination of our simulations and datasets, we have provided experimental data, models, and simulations on the public SimTK repository: https://simtk.org/projects/kuo-head-neck

**Methods**
*OpenSim Musculoskeletal Model*
The musculoskeletal model was developed in OpenSim and is derived from a previous head and neck model used to quantify moments of the neck muscles (12, 13). Details of the model are provided in Supplementary Information A and B. Our main contribution was the addition of passive ligaments in the cervical spine (this included the Annulus Fribosus fibers of the intervertebral disc). We included 80 individual sections representing 11 ligament groups (described in Supplementary Information B). The constitutive material model for the ligaments consisted of a force-length ($f^{L}_{lig}$) and force-velocity ($f^{v}_{lig}$) relationship as a function of the element length ($l^{lig}$) and element loading rate ($v^{lig}$) respectively (equation 1).

$$F_{ligament} = f^{L}_{lig}(l^{lig}) f^{v}_{lig}(v^{lig}) \qquad \text{eq. 1}$$

The ligament force-length relationship has been extensively described in literature and consists of a toe region of relatively low force production, a linear region where ligaments exhibit elastic behavior, and a yield region where the ligament begins to mechanically fail (20, 26–28). We represent ligaments here with a piece-wise linear function representing the toe and linear regions and parameterized by a toe modulus ($E_{toe}$), a toe strain ($\epsilon_{toe}$), a linear modulus ($E_{lin}$), ligament rest length ($l_0$), and ligament cross-sectional area ($A^{lig}$, equations 2-3).

$$\epsilon^{lig} = \frac{l^{lig} - l_0}{l_0} \qquad \text{eq. 2}$$

$$f^{L}_{lig}(l^{lig}) = \begin{cases} 0 & \epsilon^{lig} < 0 \\ A^{lig}\epsilon^{lig}E_{toe} & \epsilon^{lig} < \epsilon_{toe} \\ A^{lig}(E_{toe}\epsilon_{toe} + E_{lin}(\epsilon^{lig} - \epsilon_{toe})) & \epsilon^{lig} > \epsilon_{toe} \end{cases}$$
$$\text{eq. 3}$$

The ligament force-velocity relationship is also particularly well studied in cervical spine ligaments as these ligaments experience high loading rates in injury scenarios (16–18). It has been suggested that beyond a representative quasi-static loading rate (we chose 10mm/s), ligament force linearly scales with a slope ($m_{rate}$) against the log loading rate (equation 4) (18, 19).

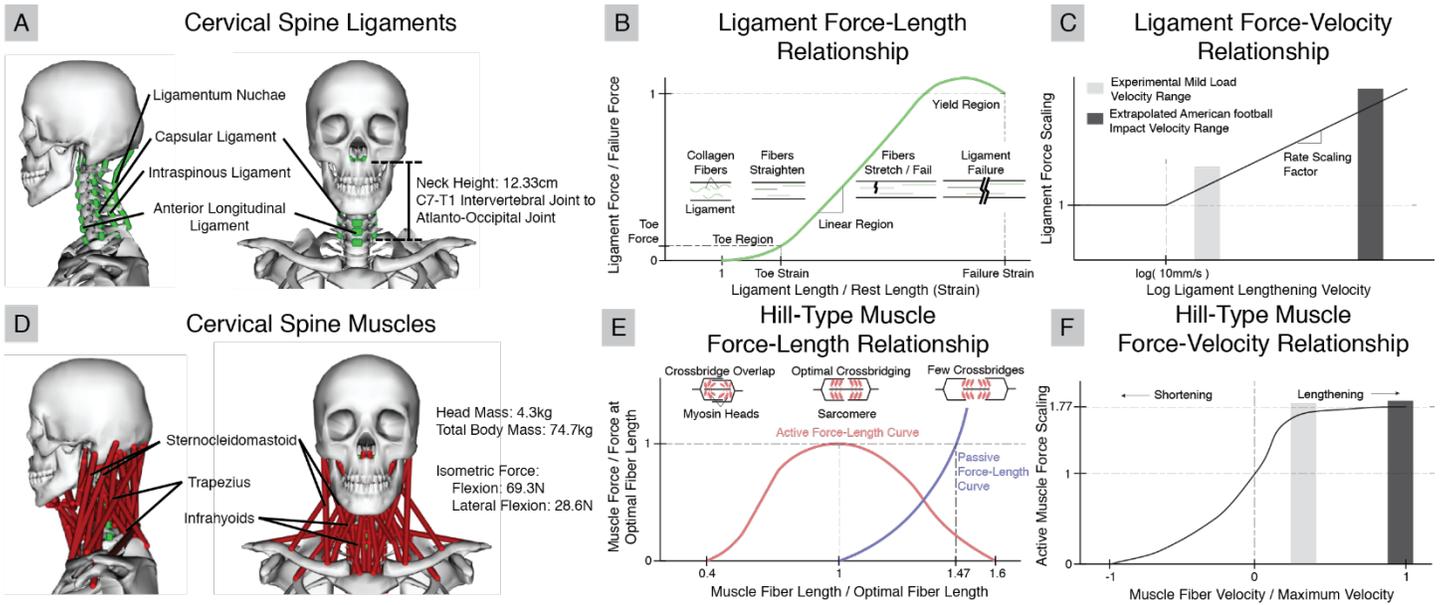

**Figure 4: OpenSim Head and Neck Impact Model:** The OpenSim musculoskeletal model is a 20 degree-of-freedom model that was based on the Mortensen 2018 model. (A) We added 80 individual cervical spine ligament sections representing 11 ligament groups were added to represent soft tissue stabilization force elements. The constitutive material model for the ligaments were represented with a (B) force-length and (C) force-velocity curve, which were identified previously in literature. The force-length relationship of ligaments have a characteristic shape corresponding to the straightening (toe region), stretching (linear region), and breaking (yield region) of collagen fibers. Our constitutive model does not include a yield region, similar to how passive tendon is modeled. (D) The model also contains 84 muscle subvolumes over 15 muscle groups, which were defined in the original Mortensen 2018 model. Muscles were represented using the Hill-Type muscle model, with a (E) force-length curve modeling myosin cross bridges in sarcomere functional units and a (F) force-velocity curve scaling force output depending on muscle fiber velocities. Ligament and muscle peak lengthening velocities are marked in (C) and (E) during experimental mild external loads and extrapolated median American football impacts. While the (E) muscle force-velocity scaling is similar in both load regimes, the (C) ligament force-velocity scaling is substantially larger during median American football impacts, resulting in greater moments produced by ligaments in the high severity regime.

$$f_{lig}^v(v^{lig}) = \begin{cases} 1 & v^{lig} < 10\frac{mm}{s} \\ 1 + m_{rate}\log(\frac{v^{lig}}{10\frac{mm}{s}}) & v^{lig} > 10\frac{mm}{s} \end{cases} \quad \text{eq. 4}$$

Parameters for each ligament at each intervertebral joint were defined and validated from previous literature (Supplementary Information B and C). Geometrical muscle and ligament attachments and bony articulations were defined visually. We additionally confirmed ligament lengths against previous literature (Supplementary Information B).

Besides the newly added ligaments, the model consists of 84 muscle subvolumes representing 15 distinct muscle groups, including hyoid muscles and multifidus muscles. These muscles were modeled as Hill-Type muscles in OpenSim (equations 5-6, Figure 4) and parameterized by optimal muscle force ($f_o^M$), activation ($a$), the active and passive force-length relationships ($f_{musc}^L$ and $f_{musc}^{PE}$ respectively) as a function of muscle length ($l^M$), and the force-velocity relationship ($f^v$) as a function of muscle velocity ($v^M$) (29). Muscle force transmission to rigid vertebral elements were parameterized by muscle fiber pennation angle ($\alpha$) and tendon force-length relationship ($f_{tendon}^L$) as a function of tendon length ($l^T$), and additionally constrained by geometrical muscle-tendon length ($l^{MT}$, equation 7). Details of the muscle model can be found in Mortensen 2018 (13).

$$F_{muscle} = f_o^M\big(af^L(l^M)f_{musc}^v(v^M) + f_{musc}^{PE}(l^M)\big) \quad \text{eq. 5}$$

$$F_{muscle}\cos(\alpha) + f_o^M f_{tendon}^L(l^T) = 0 \quad \text{eq. 6}$$

$$l^{MT} = l^M\cos(\alpha) + l^T \quad \text{eq. 7}$$

Seven cervical vertebrae (C7-C1) and the skull were represented with rigid elements. Cervical vertebrae C7 through C2 articulated with respect to the inferior vertebrae with a three degree-of-freedom rotational joint. The most inferior C7 cervical vertebrae articulated with respect to the torso with a three degree-of-freedom rotational joint. The torso was subsequently fixed to the inertial frame, though could be adjusted to articulate with full 6 degree-of-freedom motion. The C1-C2 (atlanto-axial) and skull-C1 (atlanto-occipital) joints are anatomically distinct and allow for substantial rotation about the inferior-superior axis and left-right axis respectively (30). Thus, we modeled these joints with single rotational degree-of-freedom joints about their respective rotational axes. In total, our model had 20 degrees of freedom.

Further details of the model are presented in Supplementary Information A.

*Simulating Experimental Head Impacts*
We collected a dataset of mild head impact loads in 10 human subjects (5 males and 5 females) published in a separate study with which to validate the model, and quantify the relative moment contributors (4). Briefly, we applied mild impacts to the head on subjects seated in a rigid back chair to restrict torso motion. Mild impacts were applied in two directions to produce planar head sagittal extension and head lateral flexion towards the non-dominant side to exercise primarily the dominant side sternocleidomastoid (SCM) muscle. Subject were instructed to either minimally activate (relax) or maximally cocontract neck muscles during impact. For each set of conditions (direction and muscle activity), subjects performed up to six trials.

Before applying mild impact loads, subjects performed isometric contraction trials in sagittal flexion and coronal lateral flexion while upright. This provided a measure of subject neck strength, which were used to scale model muscle strengths in OpenSim. To estimate baseline OpenSim isometric strength, we added a constraint to the skull and measured the constraint force when sagittal flexion or coronal lateral flexion muscles were maximally activated (69.3N and 28.6N isometric force in sagittal flexion and coronal lateral flexion respectively). Subject mass and estimated neck lengths from video (defined as the distance from the midpoint between the shoulders approximating the C7-T1 intervertebral joint to the atlanto-occipital joint) were also used to scale the mass and height of the OpenSim model.

Mild impact loads were delivered through the head mass center via a wrestling headgear that was attached to a load plate (2kg) upon which an impact plate (3kg) was dropped from a height of 1m. Loads were measured with an in-line tension meter (TLL-500) measuring at 1500Hz, and triggered data collection when a 50N threshold was exceeded. The tension meter collected 80ms pre-trigger and 720ms post-trigger. We averaged impact force time histories over the six trials for each subject and each set of conditions. Average impact forces were used to apply external loads in OpenSim simulations.

We measured dominant side SCM muscle activity using a custom electromyograph (EMG). Muscle activity in isometric trials was treated as the maximal activation. We previously reported average relaxed and cocontracted SCM muscle activity was 22.7±1.6% and 79.0±9.5% (mean ± standard error) of isometric muscle activity (4). Because we only measured SCM muscle activity, we had to estimate activity in the remaining neck muscles. Relaxed OpenSim neck muscle activations were found by minimizing the sum of all muscle activations while maintaining head and neck balance under gravity and with the constraint that SCM activation was 20%. Similarly, cocontracted OpenSim neck muscle activations were found by maximizing the sum of all muscle activations while maintaining head and neck balance under gravity with the constraint that SCM activation was 80%.

For each subject-scaled OpenSim model, we ran a forward dynamics simulation using relaxed or cocontracted muscle activations and applying averaged external loads (31). Four sets of conditions (sagittal extension with relaxed neck muscles, sagittal extension with cocontracted neck muscles, coronal lateral flexion with relaxed neck muscles, and coronal lateral flexion with cocontracted neck muscles) were tested for each subject, generating 40 forward dynamics simulations. For each simulation, the torso was fixed to ground to represent the rigid back chair. In sagittal extension, the symmetry of the head and neck maintained planar motion. However, in coronal lateral flexion, we had to restrict out-of-plane rotations to maintain planar motion. Detailed analysis of the simulations against the experiments is presented in Supplementary Information D.

*Moment and Angular Impulse Analysis*
After validating simulations, we computed the moments applied to the head from muscle activity, passive muscle, ligaments, external load, and gravity during the simulated experimental mild impacts. As these were all linear forces, we needed to define a moment arm for each linear force. Because the cervical spine has many degrees of freedom and several muscles cross multiple intervertebral joints, we defined the moment arm ($ma$) as the ratio between the change in length of a linear force ($dl$) and the change in rotation of the head ($d\theta$, equation 10) (12, 32).

$$ma = \frac{dl}{d\theta} \qquad \text{eq. 10}$$

The moment applied by the linear force to the head is then simply the force provided by the linear force scaled by the moment arm. Moments were first computed for each linear force element, compounded by muscle or ligament group (e.g. Hyoids), and finally aggregated by type (impact load, gravity, active muscle, passive muscle, and passive soft tissue).

To further define contributions to head stabilization, we integrated the moments to obtain angular impulses. We noted that for our controlled experimental impact, there was a distinct acceleration and deceleration phase wherein the head was set into motion (acceleration) and returned to rest (deceleration). These phases were defined using the simulated angular acceleration traces, and moments were integrated in each phase to define the angular impulse over each phase. Angular impulses were normalized by the total external load angular impulse.

*Extrapolated Median Football Impact*
Finally, we extrapolated this analysis to a more relevant American football head impact scenario. To simulate American football head impacts, we generated an extrapolated force profile that produced head kinematics similar to those previously measured from the field (14, 15). We applied a 15ms half-sine force pulse with peak 2000N to the head mass center resulting in sagittal extension and coronal lateral flexion for each of the 10 subject specific OpenSim models (Figure 3). In addition, we used the cocontracted muscle activation profile and computed relative moment and angular impulse contributions in the acceleration and deceleration phases.

**Acknowledgements**
We would like to acknowledge our funding sources, Office of Naval Research Grant N00014-16-1-2949, the Stanford Bio-X Graduate Research Fellowship Program, and the National Science Fellowship Graduate Science Fellowship. We would also like to acknowledge Jon Mortensen and Andrew Merryweather for their help with the OpenSim musculoskeletal model.

**Ethics**
Our results are based on experimental data collected in a human subject study. The lead author obtained written consent from our subjects to participate in our study, which was approved by the Stanford Internal Review Board (IRB: 36466).

**Data, Code, and Materials**
Experimental data and simulation results are posted in a public repository through the OpenSim SimTK site: https://simtk.org/projects/kuo-head-neck

**Competing Interests**
The authors have no competing interests to declare.

**Author's Contributions**
CK was responsible for designing the study, collecting experimental data, developing the model, running simulations, analyzing data, and writing the final manuscript. JS helped develop the ligament model, run simulations, and analyze data. MF helped collect experimental data, analyze data, and write the final manuscript. IY helped develop the ligament model, run simulations, and analyze data. RH helped run simulations and analyze data. Finally, DC helped with study design and writing the manuscript.

**Supplementary Information Text**

This supplementary information contains details of the OpenSim model and validation. In Supplementary Information A, we first discuss updates to the original musculoskeletal model developed by Mortensen et. al. (13). We then discuss our ligament model and the ligament properties, which were taken from previous experimental and modeling papers in Supplementary Information B. We then discuss validation of the ligaments in intersegmental simulations of the OpenSim model (Supplementary Information C) and the evaluation of the full OpenSim model against experimental mild impacts (Supplemental Information D).

**Supplementary Information A: OpenSim Musculoskeletal Model**

**Model Musculature.** The first OpenSim head and neck model was developed by Vasavada and subsequently improved by Mortensen to explore moment contributions of the neck muscles (12, 13). The Mortensen extended the original model by including the multifidus and hyoid muscles. The hyoid muscles, which all have attachments to the hyoid bone were instead attached directly to the vertebrae or skull. The hyoid bone has a complicated kinematic relationship with the cervical spine and the skull that is not easily modeled. Furthermore, we included a wrapping surface for the sternocleidomastoid subvolumes that represented the intersection of the sternocleidomastoid with the intermediate tendon of the omohyoid. As in the Mortensen model, we only modeled the superior belly of the omohyoid. The baseline model is available through SimTK: https://simtk.org/projects/kuo_head_neck.

**Model Vertebrae.** The seven vertebrae (C7 through C1) of the cervical spine, the skull (C0), and the torso (thoracic vertebra T1 and below) were all represented by rigid body elements. The inertial properties of the cervical vertebrae were taken from (33) and include the inertial properties of the vertebrae and surrounding tissue. The skull inertial properties were taken from Yoganandan 2009 (34), which matched our previous work with this data (4). Inertial properties are presented in Table A1.

**Model Kinematics.** In the original model, adjacent cervical vertebrae were joined using a three degree-of-freedom rotational joint. However, intervertebral rotations were constrained to three independent generalized coordinates resulting in smooth cervical spine curvatures throughout the range of motion. For our study, we note that previous investigations have observed different cervical spine curvature (35–37) or buckling (38–40) patterns during whiplash impacts and axial loads respectively, and thus we believed imposing cervical spine constraints would not be valid. Thus, as is stated in the manuscript, we removed kinematic constraints between cervical vertebrae C2-C7 and at the C7-T1 intervertebral joint and we represented upper cervical vertebral joint C2-C1 and C1-C0 with single degree-of-freedom rotational joints in primary axes of rotation (axial left/right rotations and sagittal flexion/extension rotations respectively). This yielded a total of 20 independent degrees-of-freedom. Cervical spine intervertebral joint degrees-of-freedom are shown in Figure S1 in Supplementary Information B.

**Table S1: Cervical Spine Inertial Properties**

| Rigid Body | Mass | Moment of Inertia Anterior-Posterior ($I_{xx}$) | Moment of Inertia Left-Right ($I_{yy}$) | Moment of Inertia Superior-Inferior ($I_{zz}$) |
|---|---|---|---|---|
| Torso* | 60.0 kg | 16,130 kg-cm$^2$ | 16,130 kg-cm$^2$ | 3,000 kg-cm$^2$ |
| [27] Cervical Vertebra C7 | 0.400 kg | 21.0 kg-cm$^2$ | 6.0 kg-cm$^2$ | 26.0 kg-cm$^2$ |
| [27] Cervical Vertebra C6 | 0.226 kg | 2.0 kg-cm$^2$ | 5.3 kgcm$^2$ | 9.7 kg-cm$^2$ |
| [27] Cervical Vertebra C5 | 0.269 kg | 6.0 kg-cm$^2$ | 4.8 kgcm$^2$ | 14.0 kg-cm$^2$ |
| [27] Cervical Verteba C4 | 0.205 kg | 1.5 kg-cm$^2$ | 3.7 kg-cm$^2$ | 5.7 kg-cm$^2$ |
| [27] Cervical Vertebra C3 | 0.156 kg | 3.3 kg-cm$^2$ | 2.6 kg-cm$^2$ | 4.3 kg-cm$^2$ |
| [27] Cervical Vertebra C2 | 0.156 kg | 3.3 kg-cm$^2$ | 2.6 kg-cm$^2$ | 4.3 kg-cm$^2$ |
| [27] Cervical Vertebra C1 | 0.156 kg | 3.3 kg-cm$^2$ | 2.6 kg-cm$^2$ | 4.3 kg-cm$^2$ |
| [28] Skull | 4.30 kg | 207.3 kg-cm$^2$ | 226.1 kg-cm$^2$ | 150.0 kg-cm$^2$ |

\* Because the torso was fixed in our simulations, the inertial properties of the torso did not affect the simulation results. These should be properly adjusted for future use.

**Supplementary Information B: Ligament Modeling and Parameters**

**Ligament Model.** The ligament model represented by equations 1-4 in the main manuscript is based on previous literature (20, 26–28). There have been several constitutive models developed to represent the non-linear force-length relationship of the ligaments; however, all of them fundamentally capture the general characteristics of the toe and linear regions, which we represent with a piece-wise linear function. While we did not model the yield region, ligament failure can be determined from simulations by identifying when yield strain or stress is achieved. However, as a caveat, previous literature has reported that failure values also have a lengthening-rate dependence (typically larger yield stresses and lower yield strains) (16, 18, 41).

**Included Ligaments.** Ligaments in the lower cervical spine (cervical vertebrae C2 through C7) are similar to those found in the thoracic spine. At each intervertebral joint, we have modeled six ligaments comprised of eight linear elements (two ligaments are represented by a pair of linear elements). These ligaments are the anterior longitudinal ligament (ALL), posterior longitudinal ligament (PLL), ligamentum flavum (LF, paired), capsular ligaments (CL, paired), interspinous ligament (ISL), and ligamentum nuchae (LN). These are shown in Figure S2. These ligaments are represented as single linear elements through the center of the ligament body (or in the case of the capsular ligament, through the center of capsule). This is similar to how muscles are modeled in the OpenSim model.

The ALL, PLL, and LN are continuous ligaments running along the length the spine. However, the geometry of these ligaments change at each spinal level, and some ligament fibers travel only between adjacent vertebrae. As such, we have chosen to represent these ligaments as individual linear elements between adjacent vertebrae. The ligamentum nuchae is unique to the cervical spine, though it is a continuation of the supraspinous ligament in the thoracic spine. Finally, the intertransverse ligament connecting the transverse processes of adjacent vertebrae are relatively small in the cervical spine, and are thus not modeled.

In addition to ligaments in the lower cervical spine, we included a pair of linear elements representing fibers of the annulus fibrosus (AF), which are part of the intervertebral discs. The AF in the cervical spine anatomically distinct from the AF in the thoracic spine. In the cervical spine, the AF is concentrated on the anterior side of the vertebrae. Furthermore, the AF fibers are well integrated with the ALL (42, 43). The orientation of the AF fibers runs at 60°-65° from the horizontal plane, and it has been suggested that these fibers provide resistance to axial rotations in addition to providing support for the disc nucleus (42, 44). We modeled our AF fibers only on the anterior side with each pair running 60° from the horizontal in opposite directions and crossing at the mid-sagittal plane.

The upper cervical spine (between C2 and the skull) are anatomically distinct from the rest of the spine. The structure of the C2 and C1 vertebrae allow for large ranges of motion in axial rotation (C2-C1 atlanto-axial joint) and flexion/extension (C1-C0 atlanto-occipital joint). Due to the large relative motion of the adjacent vertebrae, there are no intervertebral discs at these levels. Furthermore, the ligaments form a much more complex structure. In our model, we have included the anterior and posterior atlanto-atlas membranes (AAA and PAA respectively), the anterior and posterior atlanto-occipital membranes (AAO and PAO respectively), capsular ligaments (CL), alar ligaments (AL), apical ligament (API), tectorial membrane (TM), and cruciate ligament (CLT).

The anterior membranes (AAA and AAO) are continuations of the ALL, while the posterior membranes (PAA and PAO) are continuations of the LF. The remaining ligaments have attachments on the dens, a protrusion of the C2 vertebrae that allows for C1 axial rotation. There are many other minor ligaments within the upper cervical spine (accessory ligaments, barkow ligament); however, we have not included them because they are relatively weak or small compared to the ligaments described here.

**Fitting Parameters.** To fit parameters associated with the constitutive model defined by equations 1-4, we briefly reviewed literature on cervical spine ligament material properties (16–18, 20, 26–28, 41–49). Unfortunately, different researchers report different values in the constitutive model. For example: Bass et. al. (49) and Shim et. al. (17) only report yield stresses and strains; Yoganandan et. al. (26) report only linear region modulus; Chazal et. al. (50) report linear region modulus and failure stresses and strains; and Mattucci et. al. (41, 45) report toe strain and toe region modulus in addition to linear region and failure values. Furthermore, several researchers report un-normalized results in terms of forces and displacements (16–18, 20).

To compile previous literature, we primarily extracted linear region modulus, toe region modulus, and toe region strain. For studies that only report failure values (stress and strain or force and displacement), we estimated modulus or stiffness as the ratio of failure stress to failure strain, or the ratio of failure force to failure displacement. We also compiled geometrical values for the ligaments, namely cross-sectional areas and rest lengths (26–28, 46). These were used to properly non-dimensionalize studies that reported material properties in terms of force and displacement.

With material properties uniformly converted to our piecewise linear model and non-dimensionalized stress and strain values, we then fit our model to compiled data. First, we had to determine the lengthening rate dependence, as studies performed material characterization at different rates. We used reported lengthening rates ($v^{lig}$) and associated linear range moduli ($E_{lin}$) to determine $m_{rate}$. One assumption in our model is that the toe range moduli ($E_{toe}$) scales with the linear range moduli ($E_{lin}$) and toe strain does not depend on lengthening rate ($\epsilon_{toe}$). While this might not be true (given that the failure strain and stress have lengthening rate dependencies), there is insufficient data from previous literature to fit these values. Furthermore, we used 10mm/s as the representative quasi-static lengthening rate as this was the minimum lengthening rate tested for several ligaments (20).

With the fit lengthening-rate dependence $m_{rate}$, we could scale high-rate experimental parameters to fit static parameters. This procedure is visualized for the Anterior Longitudinal Ligament (ALL) segment in the intervertebral joint C7-C6 as an example in Figure S3. Fit parameters for each ligament at each intervertebral level are given in Tables S2 and S3.

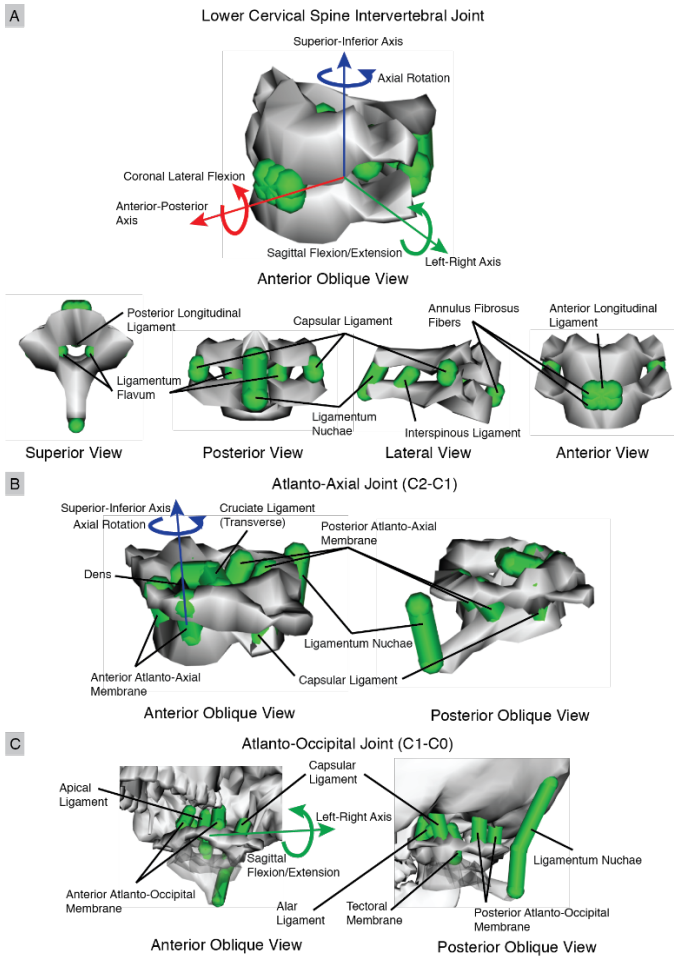

**Figure S2: Cervical Spine Ligaments and Kinematics:** (A) Lower cervical spine is represented by intervertebral joints T1-C7 through C3-C2. These joints are modeled with three rotational degrees of freedom about the anatomical axes and taken from Mortensen 2018 (13). There are six ligaments represented by eight linear elements, as well as the annulus fibrosus of the disc represented by two linear elements at 60° from the horizontal. The upper cervical spine, represented by the (B) atlanto-axial (C2-C1) joint and the (C) atlanto-occipital (C1-C0) joint are anatomically unique compared to the lower cervical spine. Of particular note, there are no intervertebral discs at this spinal level. The C2-C1 has significant compliance in axial rotation about the superior-inferior axis, with vertebrae C1 rotating about the dens of the C2 vertebra. The C1-C0 has significant compliance in sagittal flexion/extension about the left-right axis, with the skull's occipital condyles articulating on the superior facets of the C1 vertebra.

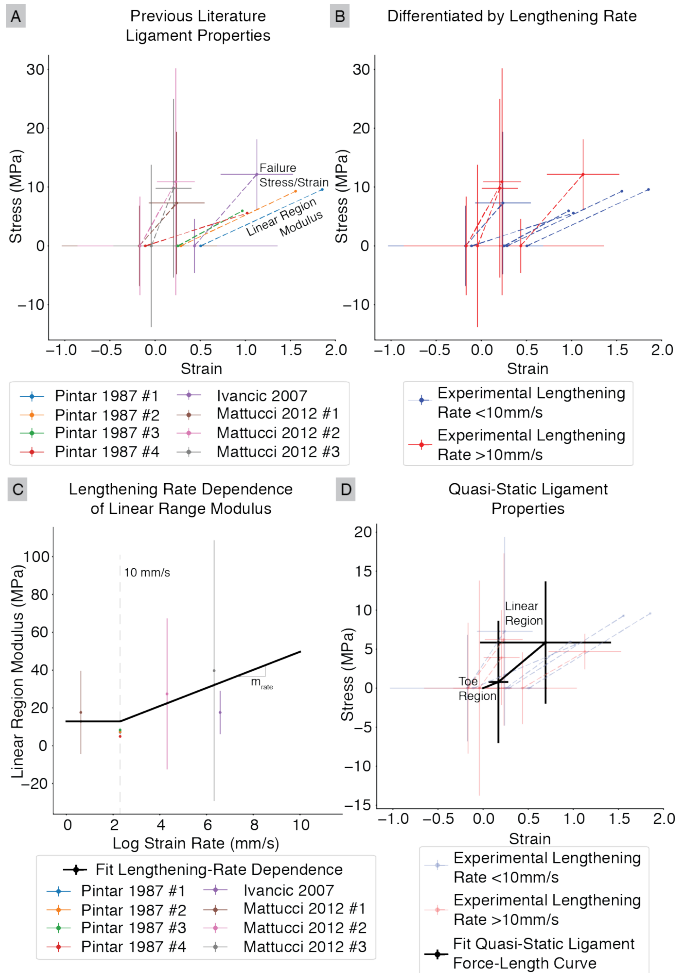

**Figure S3: Fitting Ligament Material Properties:** (A) Ligament material properties from previous literature were extracted. We primarily computed the linear region modulus (MPa), which in many reports were found by taking the ratio between the failure stress and failure strain. (B) Cervical spine ligaments are known to have a strong lengthening-rate relationship, and many studies report ligament material properties at high rates. For our study, we considered lengthening rates below 10mm/s as quasi-static, as this was the minimum lengthening rate for several ligaments previously studied. When we differentiate the previously reported stress-strain curves by lengthening rate, we observe that studies using the highest lengthening rates had larger linear region moduli. (C) Fitting a linear relationship for the linear region modulus against the log of the lengthening rate, we can solve for $m_{rate}$ from equation 7. (D) Finally, we scaled high lengthening-rate properties and fit quasi-static material properties for the ligaments.

## Table S2: Cervical Spine Ligament Material Properties (T1-C4)

### Intervertebral Joint T1-C7

| Ligament | Rest Length ($l_0$) | Cross-Sectional Area ($A^{lig}$) | Toe Strain ($\varepsilon_{toe}$) | Toe Modulus ($E_{toe}$) | Linear Modulus ($E_{lin}$) | Lengthening Rate Scaling ($m_{rate}$) |
|---|---|---|---|---|---|---|
| Anterior Longitudinal Ligament | 4.085 mm | 12.3 mm² | 27% | 4.26 MPa | 8.52 MPa | 1.33 |
| Posterior Longitudinal Ligament | 4.085 mm | 15.6 mm² | 17% | 5.02 MPa | 10.04 MPa | 1.26 |
| Ligamentum Flavum | 9.420 mm | 21.3 mm² | 26% | 3.65 MPa | 7.85 MPa | 1.17 |
| Capsular Ligament | 4.329 mm | 55.5 mm² | 17% | 0.99 MPa | 1.98 MPa | 1.17 |
| Interspinous Ligament | 9.795 mm | 23.9 mm² | 13% | 1.26 MPa | 2.53 MPa | 1.29 |
| Ligamentum Nuchae | 21.586 mm | 10.0 mm² | 5% | 25 MPa | 50 MPa | 1.20 |
| Annulus Fibrosus Fibers | 8.983 mm | 100.0 mm² | 5% | 25 MPa | 50 MPa | 1.50 |

### Intervertebral Joint C7-C6

| Ligament | Rest Length ($l_0$) | Cross-Sectional Area ($A^{lig}$) | Toe Strain ($\varepsilon_{toe}$) | Toe Modulus ($E_{toe}$) | Linear Modulus ($E_{lin}$) | Lengthening Rate Scaling ($m_{rate}$) |
|---|---|---|---|---|---|---|
| Anterior Longitudinal Ligament | 4.056 mm | 11.75 mm² | 27% | 4.23 MPa | 8.45 MPa | 1.38 |
| Posterior Longitudinal Ligament | 4.056 mm | 14.65 mm² | 17% | 4.84 MPa | 9.68 MPa | 1.37 |
| Ligamentum Flavum | 9.145 mm | 20.70 mm² | 26% | 3.44 MPa | 6.88 MPa | 1.22 |
| Capsular Ligament | 4.087 mm | 52.39 mm² | 17% | 1.04 MPa | 2.08 MPa | 1.15 |
| Interspinous Ligament | 11.081 mm | 14.54 mm² | 13% | 1.73 MPa | 3.46 MPa | 1.28 |
| Ligamentum Nuchae | 13.762 mm | 10.0 mm² | 5% | 10 MPa | 20 MPa | 1.20 |
| Annulus Fibrosus Fibers | 8.970 mm | 100.0 mm² | 5% | 25 MPa | 50 MPa | 1.50 |

### Intervertebral Joint C6-C5

| Ligament | Rest Length ($l_0$) | Cross-Sectional Area ($A^{lig}$) | Toe Strain ($\varepsilon_{toe}$) | Toe Modulus ($E_{toe}$) | Linear Modulus ($E_{lin}$) | Lengthening Rate Scaling ($m_{rate}$) |
|---|---|---|---|---|---|---|
| Anterior Longitudinal Ligament | 4.074 mm | 10.76 mm² | 27% | 4.70 MPa | 9.40 MPa | 1.25 |
| Posterior Longitudinal Ligament | 3.857 mm | 13.07 mm² | 17% | 6.85 MPa | 13.71 MPa | 1.23 |
| Ligamentum Flavum | 7.827 mm | 17.83 mm² | 26% | 1.99 MPa | 3.98 MPa | 1.36 |
| Capsular Ligament | 3.839 mm | 47.19 mm² | 17% | 0.77 MPa | 1.53 MPa | 1.23 |
| Interspinous Ligament | 8.714 mm | 20.65 mm² | 13% | 1.26 MPa | 2.52 MPa | 1.28 |
| Ligamentum Nuchae | 15.687 mm | 10.0 mm² | 5% | 10 MPa | 20 MPa | 1.20 |
| Annulus Fibrosus Fibers | 8.978 mm | 40.0 mm² | 5% | 25 MPa | 50 MPa | 1.50 |

### Intervertebral Joint C5-C4

| Ligament | Rest Length ($l_0$) | Cross-Sectional Area ($A^{lig}$) | Toe Strain ($\varepsilon_{toe}$) | Toe Modulus ($E_{toe}$) | Linear Modulus ($E_{lin}$) | Lengthening Rate Scaling ($m_{rate}$) |
|---|---|---|---|---|---|---|
| Anterior Longitudinal Ligament | 4.268 mm | 10.29 mm² | 27% | 4.22 MPa | 8.45 MPa | 1.43 |
| Posterior Longitudinal Ligament | 3.679 mm | 11.33 mm² | 17% | 7.00 MPa | 14.00 MPa | 1.35 |
| Ligamentum Flavum | 7.604 mm | 17.33 mm² | 26% | 1.85 MPa | 3.69 MPa | 1.40 |
| Capsular Ligament | 4.662 mm | 57.30 mm² | 17% | 0.80 MPa | 1.59 MPa | 1.22 |
| Interspinous Ligament | 10.351 mm | 11.22 mm² | 13% | 2.90 MPa | 5.80 MPa | 1.12 |
| Ligamentum Nuchae | 12.572 mm | 10.0 mm² | 5% | 10 MPa | 20 MPa | 1.20 |
| Annulus Fibrosus Fibers | 9.067 mm | 40.0 mm² | 5% | 25 MPa | 50 MPa | 1.50 |

* Ligamentum Nuchae and Annulus Fibrosus Fiber properties were estimated from other modeling studiess rather than experimental studies or matched to the Interspinous Ligament or Anterior Longitudinal Ligament respectively.

# Table S3: Cervical Spine Ligament Material Properties (C4-Skull)

### Intervertebral Joint C4-C3

| Ligament | Rest Length ($l_o$) | Cross-Sectional Area ($A^{lig}$) | Toe Strain ($\varepsilon_{toe}$) | Toe Modulus ($E_{toe}$) | Linear Modulus ($E_{lin}$) | Lengthening Rate Scaling ($m_{rate}$) |
|---|---|---|---|---|---|---|
| Anterior Longitudinal Ligament | 3.612 mm | 10.40 mm² | 27% | 3.98 MPa | 7.96 MPa | 1.37 |
| Posterior Longitudinal Ligament | 4.159 mm | 10.56 mm² | 17% | 6.42 MPa | 12.83 MPa | 1.30 |
| Ligamentum Flavum | 8.106 mm | 21.35 mm² | 26% | 3.20 MPa | 6.40 MPa | 1.15 |
| Capsular Ligament | 4.111 mm | 45.77 mm² | 17% | 1.32 MPa | 2.64 MPa | 1.16 |
| Interspinous Ligament | 7.829 mm | 19.30 mm² | 13% | 1.71 MPa | 3.42 MPa | 1.32 |
| Ligamentum Nuchae | 13.509 mm | 10.0 mm² | 5% | 25 MPa | 50 MPa | 1.20 |
| Annulus Fibrosus Fibers | 8.778 mm | 30.0 mm² | 5% | 25 MPa | 50 MPa | 1.50 |

### Intervertebral Joint C3-C2

| Ligament | Rest Length ($l_o$) | Cross-Sectional Area ($A^{lig}$) | Toe Strain ($\varepsilon_{toe}$) | Toe Modulus ($E_{toe}$) | Linear Modulus ($E_{lin}$) | Lengthening Rate Scaling ($m_{rate}$) |
|---|---|---|---|---|---|---|
| Anterior Longitudinal Ligament | 4.106 mm | 11.28 mm² | 27% | 4.61 MPa | 9.22 MPa | 1.35 |
| Posterior Longitudinal Ligament | 3.942 mm | 9.72 mm² | 17% | 8.78 MPa | 16.56 MPa | 1.22 |
| Ligamentum Flavum | 7.599 mm | 20.01 mm² | 26% | 2.18 MPa | 4.35 MPa | 1.30 |
| Capsular Ligament | 4.116 mm | 45.82 mm² | 17% | 1.21 MPa | 2.42 MPa | 1.19 |
| Interspinous Ligament | 9.797 mm | 9.77 mm² | 13% | 2.79 MPa | 5.47 MPa | 1.19 |
| Ligamentum Nuchae | 14.090 mm | 10.0 mm² | 5% | 10 MPa | 20 MPa | 1.20 |
| Annulus Fibrosus Fibers | 8.992 mm | 20.0 mm² | 5% | 25 MPa | 50 MPa | 1.50 |

### Intervertebral Joint C2-C1 (Atlanto-Axial)

| Ligament | Rest Length ($l_o$) | Cross-Sectional Area ($A^{lig}$) | Toe Strain ($\varepsilon_{toe}$) | Toe Modulus ($E_{toe}$) | Linear Modulus ($E_{lin}$) | Lengthening Rate Scaling ($m_{rate}$) |
|---|---|---|---|---|---|---|
| Anterior Atlanto-Axial Membrane | 11.812 mm | 23.84 mm² | 7.6% | 5.21 MPa | 10.41 MPa | 1.30 |
| Posterior Atlanto-Axial Membrane | 15.147 mm | 10.38 mm² | 8.6% | 10.83 MPa | 21.65 MPa | 1.30 |
| Capsular Ligament | 14.936 mm | 39.83 mm² | 7.8% | 5.87 MPa | 10.74 MPa | 1.73 |
| Ligamentum Nuchae | 22.963 mm | 10.0 mm² | 5% | 10 MPa | 20 MPa | 1.20 |
| Cruciate Ligament (Transverse) | 20.309 mm | 18.57 mm² | 11% | 26.92 MPa | 53.83 MPa | 1.31 |

### Intervertebral Joint C1-C0 (Atlanto-Occipital)

| Ligament | Rest Length ($l_o$) | Cross-Sectional Area ($A^{lig}$) | Toe Strain ($\varepsilon_{toe}$) | Toe Modulus ($E_{toe}$) | Linear Modulus ($E_{lin}$) | Lengthening Rate Scaling ($m_{rate}$) |
|---|---|---|---|---|---|---|
| Anterior Atlanto-Occipital Membrane | 12.735 mm | 44.30 mm² | 12% | 0.64 MPa | 1.28 MPa | 2.56 |
| Posterior Atlanto-Occipital Membrane | 13.600 mm | 47.72 mm² | 8.4% | 0.39 MPa | 0.78 MPa | 2.85 |
| Capsular Ligament | 19.088 mm | 38.17 mm² | 7.1% | 23.64 MPa | 47.28 MPa | 1.38 |
| Ligamentum Nuchae | 40.050 mm | 10.0 mm² | 5% | 10 MPa | 20 MPa | 1.20 |
| Alar Ligament** | 10.146 mm | 20.96 mm² | 5% | 8.05 MPa | 16.10 MPa | 1.13 |
| Apical Ligament** | 13.925 mm | 19.03 mm² | 7.8% | 10.97 MPa | 20.94 MPa | 1.75 |
| Tectorial Membrane** | 18.520 mm | 32.32 mm² | 7.8% | 20.99 MPa | 40.97 MPa | 1.75 |

\* Ligamentum Nuchae and Annulus Fibrosus Fiber properties were estimated from other modeling studies rather than experimental studies or matched to the Interspinous Ligament or Anterior Longitudinal Ligament respectively.
\*\* These ligaments cross from C2 to the skull via the Dens

## Supplementary Information C: Cervical Spine Ligament Validation

To validate the cervical spine ligaments, we chose to compare intervertebral segment moment-deflection behavior against previously published experiments (23, 51–53). In these experiments, researchers applied pure moments to functional intervertebral segments (51–53) or whole cervical spine specimens (23) and measured the resulting rotational deflection. For all experiments, muscles were typically excised, leaving the vertebrae, ligaments, and discs. Soft tissue (ligaments and discs) are thus responsible for moment-deflection characteristics. We felt these experiments were ideal for validation because moments depend on both the material and geometrical properties of the soft tissue.

To validate our model, we created matching functional intervertebral models with only vertebrae and associated ligaments. For the upper cervical spine, we created one model with cervical vertebrae C2 through the skull to match Goel et. al. (51). Note, because we chose to model the two joints between cervical vertebrae C2 through the skull with single orthogonal degrees of freedom, their motions could be considered independent. According to Nightingale et. al. (52), the ligamentum nuchae was also excised as this ligament is an insertion point for many posterior muscles. Thus, we also removed the ligamentum nuchae from our functional intervertebral models.

With the models, we prescribed rotations similar to those in previous literature. Lower cervical spine intervertebral segments were rotated from -10˚ to 10˚ in coronal lateral flexion about the anterior-posterior axis, sagittal flexion/extension about the left-right axis, and axial rotation about the superior-inferior axis. The upper cervical spine model representing C2 through skull was rotated from -20˚ to 20˚ in sagittal flexion/extension about the left-right axis and axial rotation about the superior-inferior axis. Note, coronal lateral flexion rotations about the anterior-posterior axis were not exercised because the C2 through skull model does not have a degree of freedom allowing for that rotational direction. Furthermore, a greater range was exercised in previous literature and in our validations due to the increased compliance in the upper cervical spine (23, 51).

Moments provided by the soft tissue during these rotations were computed using the same method reported in the main manuscript (equation 10). Briefly, this method computes a moment arm for linear force elements using the ratio between the change in rotation angle and the change in length of the linear element (32). The moment is then the moment arm scaled by the force produced. Figures S3 and S4 show the simulated moment-deflection curves for functional intervertebral units overlaid on previous experimental data. Furthermore, moment contributions from individual ligament linear elements are included.

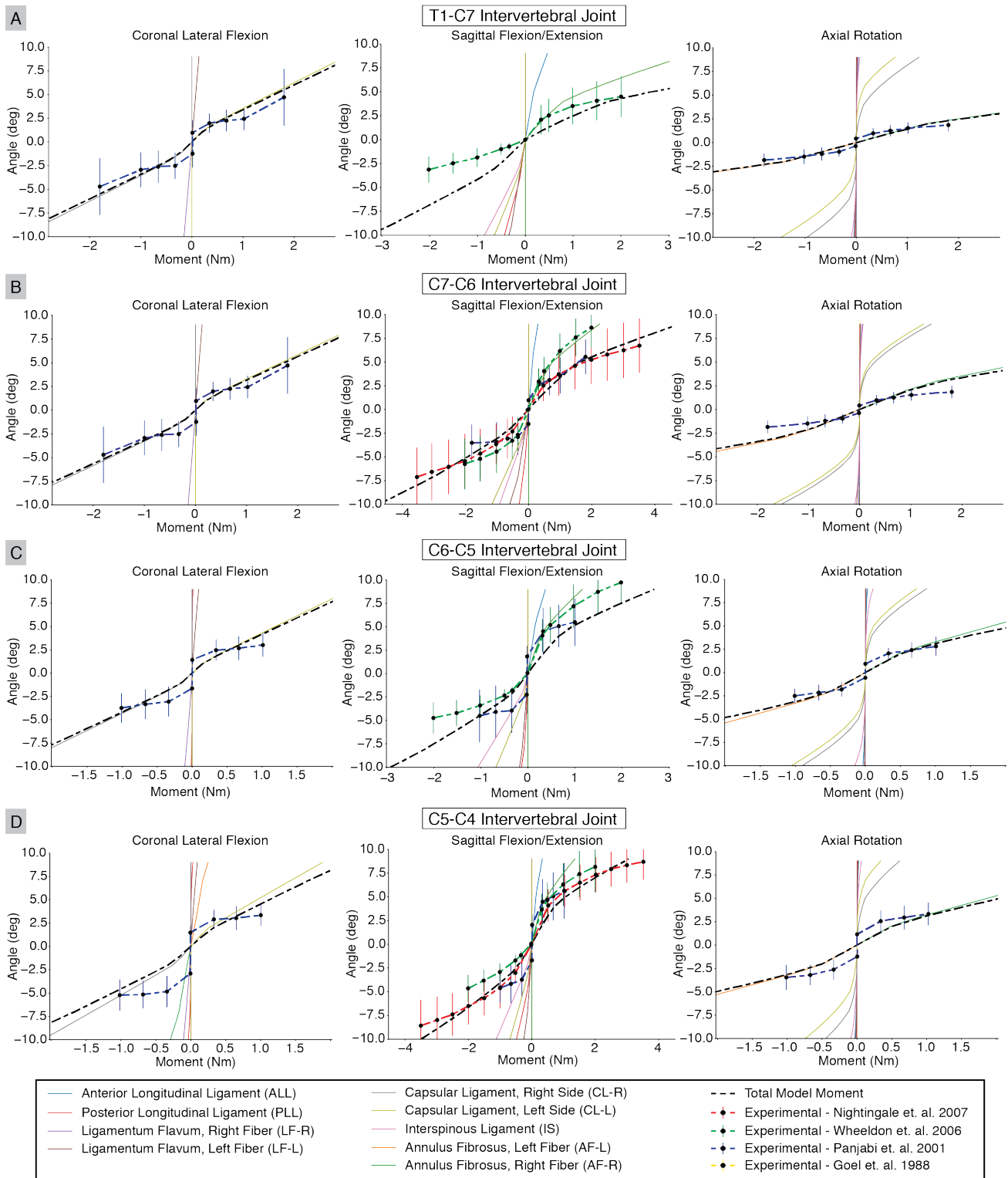

**Figure S3: Intersegmental Validation T1-C4:** We simulated intervertebral moment-deflection curves to validate our model against experimental data (23, 51–53). The intervertebral functional units contain adjacent vertebrae as well as the ligaments and intervertebral disc. The model ligamentum nuchae is removed to match experimental conditions in (52).

**Figure S4: Intersegmental Validation C4-Skull:** We simulated intervertebral moment-deflection curves to validate our model against experimental data (23, 51–53). The intervertebral functional units contain adjacent vertebrae as well as the ligaments and intervertebral disc. The model ligamentum nuchae is removed to match experimental conditions in (52). We modeled the full upper cervical spine from C2 through the skull as a single functional unit.

# Supplementary Information D: Full Model Evaluation against Experimental Mild Loads

**Simulation Analysis.** To evaluate the OpenSim simulation of the experimental mild impact loads, we first show an example of the head center of mass trajectory with respect to the torso in each of four conditions (sagittal plane extension with relaxed neck muscles, sagittal plane extension with cocontracted neck muscles, coronal plane lateral flexion with relaxed neck muscles, and coronal plane lateral flexion with cocontracted neck muscles) in one subject (Figure S5). In the experiments, head center of mass trajectory was determined through high speed video tracking, as described in previous work. In the simulation, head center of mass trajectory was extracted directly using forward kinematics. In both cases, the zero position and zero angle were defined as the position and orientation at time zero.

Angular velocity, angular acceleration, and linear acceleration kinematics are more commonly assessed in head impact analysis and thus, we also evaluated OpenSim simulation errors in these kinematic measures. Planar angular velocities and angular acceleration were about the left-right axis in sagittal extension, and about the anterior-posterior axis in coronal lateral flexion. Planar linear accelerations were evaluated at the head center of mass along the anterior-posterior axis in sagittal extension and along the left-right axis in coronal lateral flexion. We did not compare planar linear accelerations along the superior-inferior axis because they were relatively small.

In our experiments, subjects were equipped with an instrumented bite-bar measuring head inertial tri-axial angular velocity and tri-axial linear acceleration at 10kHz with 100ms pre-trigger and 500ms post-trigger. Sensor axes were aligned with anatomical axes (anterior-posterior, left-right, and superior-inferior), and data were filtered with a 50Hz lowpass 4$^{th}$ order butterworth filter. We differentiated angular velocities using a 5-point stencil to obtain angular accelerations. As with the external forces, we average the kinematics time histories over the six trials for each subject and each set of conditions.

Simulated kinematics were extracted directly from the simulations. Angular velocities and angular accelerations were of the skull with respect to the laboratory frame (or the torso frame as the torso was fixed to the laboratory frame). Linear accelerations were of the skull center of mass with respect to the laboratory frame.

We computed three metrics to compare the experimental and simulated kinematics. Before computing these metrics, we first resampled experimental kinematics to 1000Hz with 50ms pre-trigger and 250ms post-trigger ($n = 300$ samples). First, we computed the Pearson's correlation coefficient between the kinematics signals using the Matlab "corr" function. Second, we computed the normalized root mean square (NRMS) error between the experimental ($x_{exp}$) and simulated ($x_{sim}$) kinematics, normalizing by the experimental range (equation S1). Finally, we computed the normalized error in kinematic range (equation S2).

$$NRMS = \frac{\sqrt{\frac{\Sigma_n(x_{exp}-x_{sim})^2}{n}}}{\max(x_{exp})-\min(x_{exp})} \qquad \text{eq. S1}$$

$$Range\ Error = \frac{[\max(x_{exp})-\min(x_{exp})]-[\max(x_{sim})-\min(x_{sim})]}{\max(x_{exp})-\min(x_{exp})} \qquad \text{eq. S2}$$

**Analysis Results.** Example traces for average experimental and simulated angular velocities, angular accelerations, and linear accelerations are shown in Figure S6. Analysis shows that simulated kinematics had correlations exceeding 80%, NRMS errors below 40%, and peak to peak errors below 50% (Figure S7).

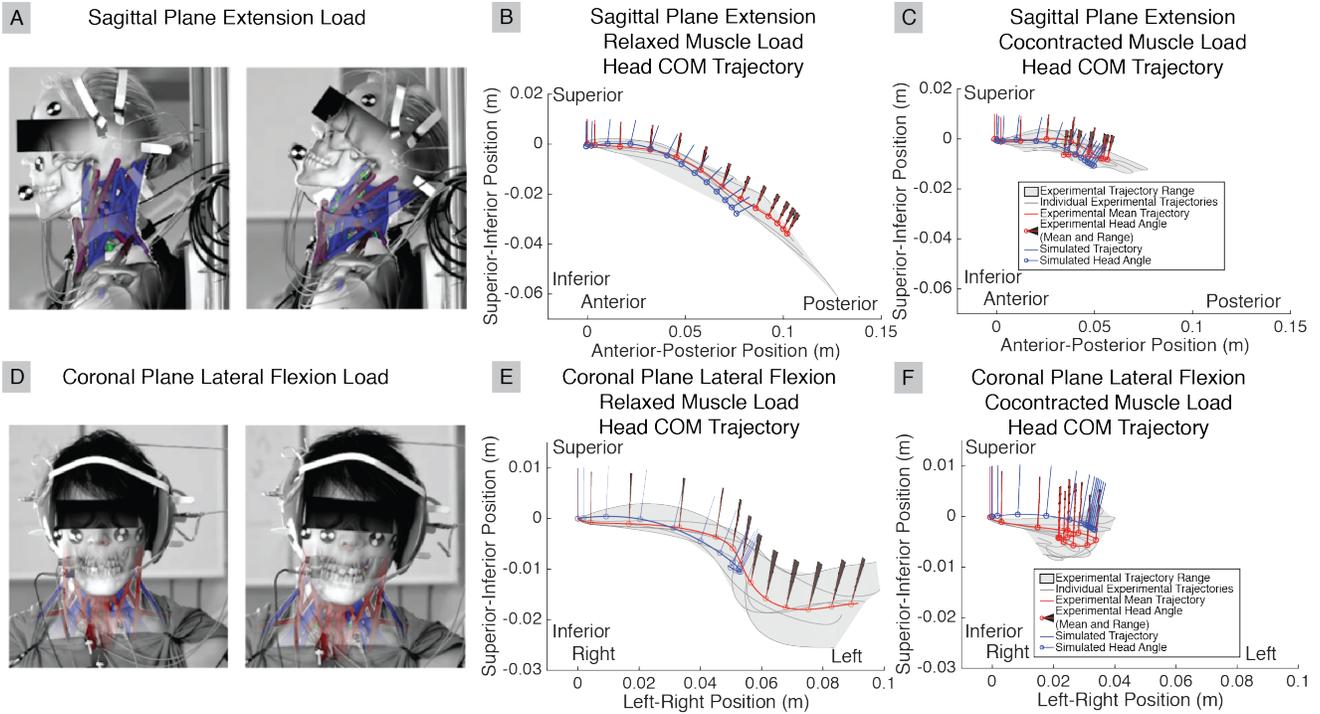

**Figure S5: Example Head Center of Mass Trajectories:** OpenSim screenshots were overlaid on experimental videos for (A) sagittal extension impacts and (D) coronal plane impacts to qualitatively demonstrate similar behavior. In the experimental videos, we tracked the trajectory of the head center of mass with respect to the torso and the head's orientation with respect to the torso. We compared simulated trajectories against experimental trajectories, with examples of (B) sagittal plane extension load with relaxed muscle activity, (C) sagittal plane extension load with cocontracted muscle activity, (E) coronal plane lateral flexion load with relaxed muscle activity, and (F) coronal plane lateral flexion load with cocontracted muscle activity shown here. In these sample traces, the simulated trajectory falls within the range of trajectories for the six experimental trials within a given condition for a single subject.

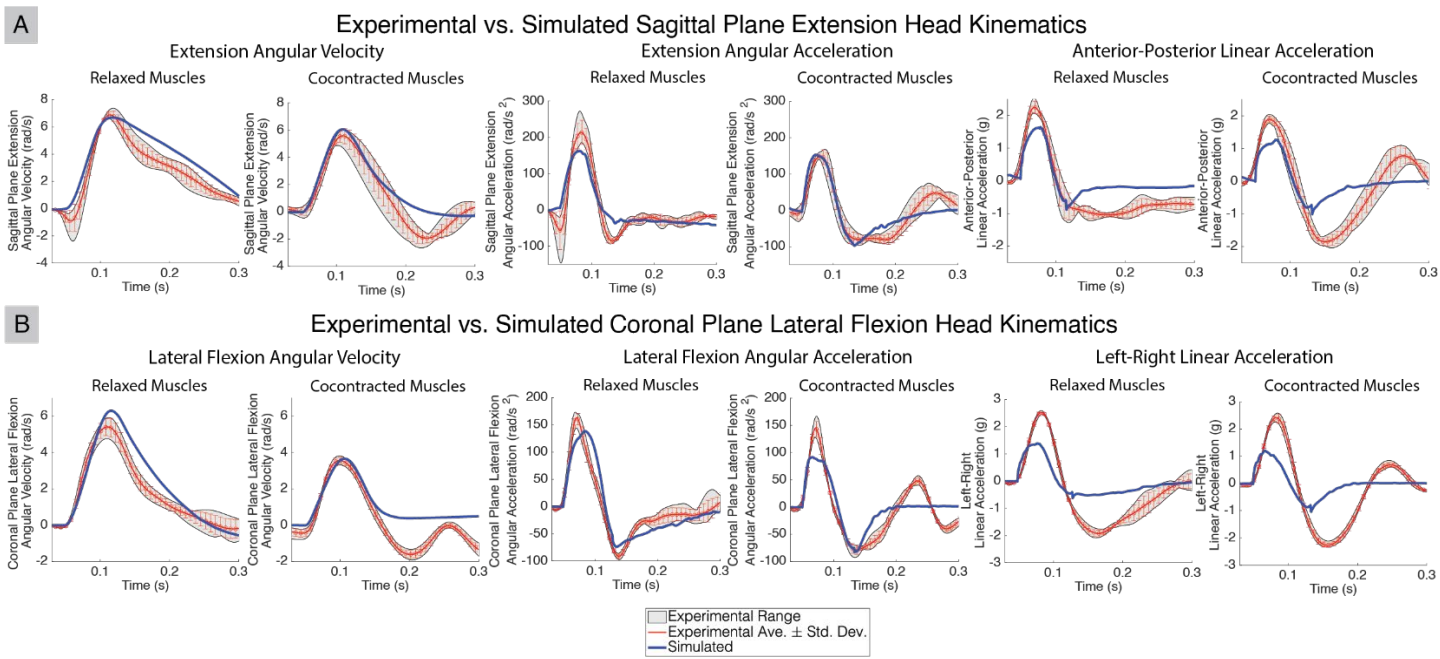

**Figure S6: Comparing Sample Experimental and Simulated Kinematics:** Sample experimental and simulated kinematics from (A) sagittal extension and (B) coronal lateral flexion also show similar behavior. Planar angular velocity, angular acceleration, and linear accelerations, are all shown for both relaxed muscle and cocontracted muscle conditions. Experimental data were aggregated over the six trials for each condition, with the minimum to maximum range, average, and standard deviation shown here.

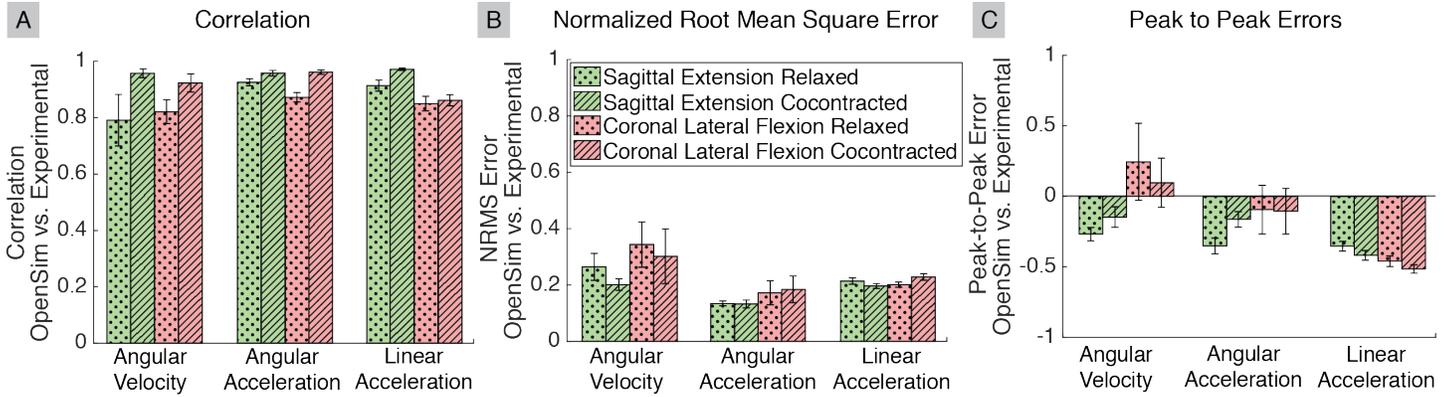

**Figure S7: Validation between Simulated OpenSim and Experimental Kinematics:** The planar angular velocity, angular acceleration, and liner acceleration at the head center of mass were computed from OpenSim simulations and taken from instrumented bite-bar data in experimental trials. (A) Correlation, (B) NRMS error, and (C) peak to peak errors between OpenSim and experimental kinematics were computed for each condition and aggregated over all 10 subjects (error bars represent standard errors). In most conditions, OpenSim and experimental kinematics had above 80% correlation, below 40% NRMS errors, and below 50% peak to peak errors.